# Haptic Information Feedback Given to Handles in Guide Dog Training

Chonghoon Park[1], Qirong Zhu[1], Shinji Tanaka[2], Yasutoshi Makino[1], and Hiroyuki Shinoda[1]

[1] *Graduate School of Frontier Sciences, University of Tokyo, Chiba, Japan*

[2] *Japan Guide Dog Association, Kanagawa, Japan*

(Email: park, e.zhu, @hapis.k.u-tokyo.ac.jp

yasutoshi_makino, hiroyuki_shinoda @k.u-tokyo.ac.jp)

**Abstract ---** In guide dog training, trainers use haptic information transmitted through the handle of the harness worn by the guide dog to understand the dog's state. They then apply appropriate force to the handle to train the dog to make correct judgments. This tactile experience can only be felt between the dog and the trainer, making it challenging to communicate the amount of force applied to others quantitatively. To solve this problem, this study proposes a method for real-time visualization of the force exerted on the handle and quantification of the handle movement through image processing, which can be applied to actual guide dog training.

## 1 Introduction

This study proposes a method to visualize and utilize force information occurring at the handle part of the harness worn by guide dogs (hereafter referred to as the "handle") to improve the efficiency of guide dog training. When guide dog trainers train guide dogs, they comprehend the surrounding situation and obstacles through visual and auditory senses, observe the behavior of the guide dog, and assess the dog's reactions obtained through the handle. They then comprehensively judge the situation based on all these factors. Therefore, the movement of the handle contains information exchanged between the trainer and the guide dog in both directions. Understanding this exchange of information is crucial in the utilization and training of guide dogs.

When considering the training of guide dog trainers, this haptic information is challenging to observe and understand from the outside when an experienced trainer tries to teach a skill to a novice trainer, making it challenging to convey the appropriate amount of force and timing. Currently, experienced trainers observe the training process externally, estimate the degree of force based on the behavior of both the novice trainer and the guide dog, and provide guidance accordingly. Alternatively, they may place their hand directly over the novice's hand on the handle to demonstrate the appropriate force to apply.

Our goal is to solve this problem by realizing a system that measures the applied force and displays it in real-time on a PC screen to evaluate force and utilize it for skill transfer quantitatively. Previous research addressing the difficulty of transmitting such sensory information includes attempts to classify and quantify guide dog behavior using Inertial Measurement Units (IMUs) [1]. Additionally, Zhu et al. have proposed a system that incorporates sensors into the handle to measure information during training [2]. Zhu's system uses accelerometers, force sensors, and depth cameras mounted on the handle to measure the handle's inclination and force simultaneously. We have previously developed a prototype system based on Zhu's work, attaching wireless force sensors to the handle to provide real-time feedback on troops during training. This allowed for quantitative analysis of training and clarification of trainers' requirements [3]. We confirmed that the rotation of the handle relative to the back and the direction of pushing and pulling are essential elements.

In this study, we measured force through sensors and attached a 360-degree camera to the trainer's shoulder to estimate the dog's body orientation from the captured images. As shown in Fig. 1, this allowed us to measure both the axial force on the handle and the relative angle to the dog's back. This approach confirmed the potential for application in a wider variety of scenarios in guide dog training that involve handle rotation.

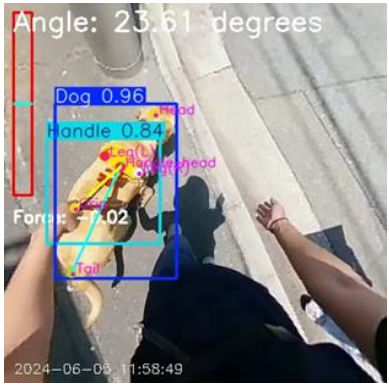

Fig.1 Example of force and angle information provided

## 2 METHOD

### 2.1 Device construction

Figures 2 and 3 show the configuration of the measurement system and its block diagram. The measurements are conducted using two devices: force measurement with a force sensor and detection of the movements of the guide dog and handle based on 360-degree camera (RICOH THETA Z1) video.

We tapped the handle several times and synchronized the results from the force sensor with the recorded video footage, aligning the camera and the force sensor data.

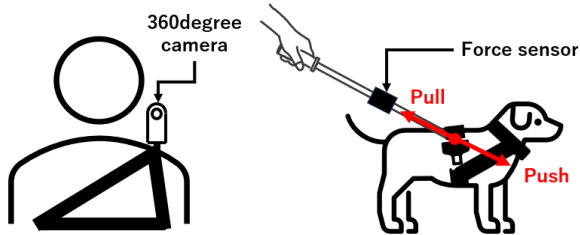

Fig.2 Configuration of system

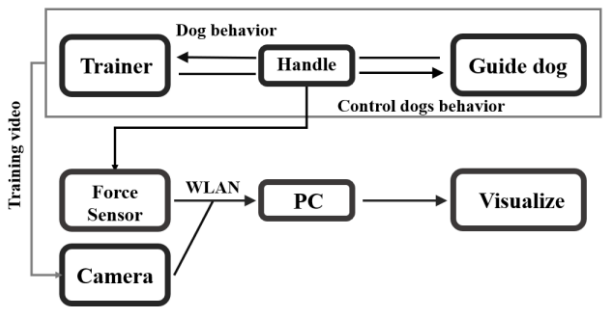

Fig.3 System block diagram

### 2.2 Detect handle and guide the dog.

In the training process, the movements of the guide dog and the handle are detected using the pose estimation model of Yolo V8, one of the deep learning algorithms. Using a 360-degree camera video recorded above the trainer's shoulder, images in an equirectangular projection format are obtained. The area in the lower front where the dog is located is converted into perspective projection images of 416*416 pixels, and 760 images are extracted. These images are labeled as shown in Fig. 4, and additional training is performed on a model pre-trained with the COCO dataset [4] to detect the movements of the guide dog and the trainer's handle.

To evaluate the accuracy of the trained model, the difference between the predicted and the manually labeled key point coordinates was calculated for 48 test images that were not used in the training. Table 1 shows the average pixel difference between the key points' actual and estimated values, and Table 2 shows the detection failure percentage for the key points.

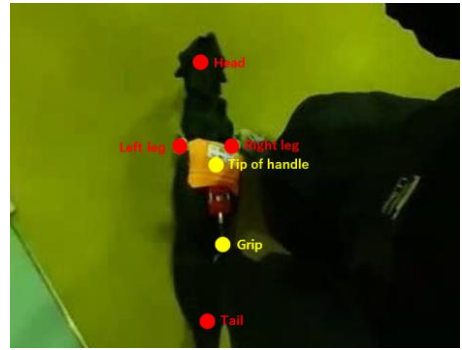

Fig.4 Labeling example

Table 1 Key Points Detection Accuracy

| subject | Left leg | Right leg | Head | Tail | handle | grip |
|---|---|---|---|---|---|---|
| x[px] | 3.2 | 3.0 | 2.9 | 11.2 | 1.6 | 2.0 |
| y[px] | 4.2 | 4.6 | 3.4 | 3.4 | 7.8 | 2.4 |

Table 2 Detection Failure Percentage

| subject | Left leg | Right leg | Head | Tail | handle | grip |
|---|---|---|---|---|---|---|
| Failure [%] | 0 | 2.08 | 2.08 | 14.5 | 0 | 2.08 |

### 2.3 Force Measurement

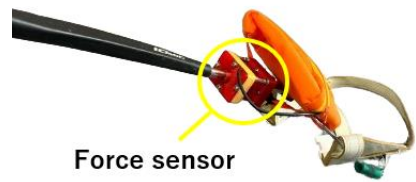

Fig.5 Handle with sensors

As shown in Fig. 5, we cut a commercially available guide dog harness handle for force measurement and installed a 6-axis force sensor (Leptrino CFS018CA201U) in the middle.

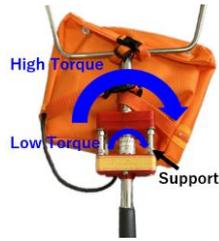

Fig.6  Support to protect force sensor

In this setup, there was a risk of sensor damage if a load higher than the specified value was applied in the direction of rotation or bending of the handle axis. To prevent this, we attached metal support materials to the force sensor to reduce twisting and bending, thereby limiting the application of loads exceeding the specified value to the handle.

## 3  Conclusion

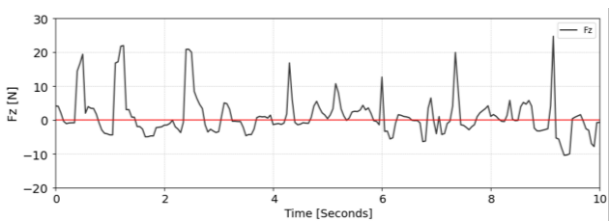

Fig.7  Force changes while walking

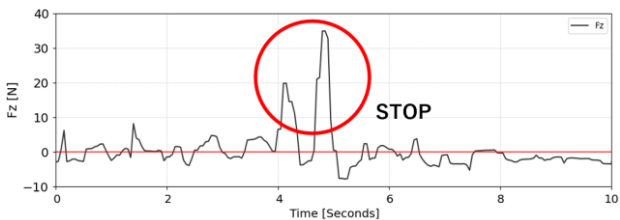

Fig.8  Force changes during a sudden stop

In this system, along with force measurement, we also conducted image-based handle movement measurements, confirming the potential for use in various scenarios.

Figures 6 and 7 show some of the actual measurement results. In Fig. 6, during walking, pulling and pushing forces occur in rhythm with the human's walking cycle. Although it varies depending on the state of walking training, approximately 20N of force is applied. In Fig. 7, when stopping, the dog stops first, and the human catches up, so a stronger pushing force is measured. This was measured as a peak of over 30N in the force sensor output.

In the demonstration, the harness was not attached to the dog. Instead, participants will wear the 360-degree camera and handle introduced in Section 2, allowing for the reproduction of real-time force feedback and handle detection.

ACKNOWLEDGMENT

This work is supported in part by 21H05301.

REFERENCES

[1] Z. Cleghern, E. Williams, S. Mealin, M. Foster, T. Holder, A. Bozkurt, and D. L. Roberts: "An IoT and Analytics Platform for Characterizing Adolescent Dogs' Suitability for Guide Work," Conference on Animal-Computer Interaction, No.1, pp. 1–6 (2019)

[2] Q. Zhu, A. Wang, S. Tanaka, Y. Matsunami, Y. Makino, and H. Shinoda: "Haptic Recording and Display System for Guide Dog," IEEE World Haptics Conference, (2023).

[3] C. Park, Q. Zhu, S. Tanaka, Y. Matsunami, Y. Makino, and H. Shinoda: "Visualizing the Force Applied to the Handle in Guide Dog Training," 40th SICE Sensing Forum, (2023).

[4] T.-Y. Lin, M. Maire, S. Belongie, J. Hays, P. Perona, D. Ramanan, P. Dollár, and C. L. Zitnick: "Microsoft COCO: Common Objects in Context," European Conference on Computer Vision (ECCV), pp. 740-755 (2014).